\documentclass[prd,epsf,aps,twocolumn]{revtex4}
\usepackage{amsmath}
\usepackage{amssymb}
\usepackage{eucal}
\usepackage[all]{xy}
\input xypic
\numberwithin{equation}{section}
\begin{document}
\title{Novel Geometrical Models of Relativistic Stars.\\
       II. Incompressible Stars and Heavy Black Dwarfs}
\author{P.~P.~Fiziev\footnote{ E-mail:\,\, fiziev@phys.uni-sofia.bg} }
\affiliation{Department of Theoretical Physics, Faculty of
Physics, Sofia University, 5 James Bourchier Boulevard,
Sofia~1164, Bulgaria.\\and\\ Joint Institute of Nuclear Research,
Dubna, Russia. }
\begin{abstract}
In a series of articles we describe a novel class of geometrical
models of relativistic stars. Our approach to the static
spherically symmetric solutions of Einstein equations is based on
a careful physical analysis of radial gauge conditions.

It turns out that there exist heavy black dwarfs: relativistic
stars with arbitrary large mass, which are to have arbitrary small
radius and arbitrary small luminosity. In the present article we
mathematically  prove this new phenomena, using a detailed
consideration of incompressible GR stars. We study the whole two
parameter family of solutions of extended TOV equations for
incompressible stars. This example is used to illustrate most of
the basic features of the new geometrical models of relativistic
stars. Comparison with newest observational data is discussed.

\noindent{PACS number(s): 04.20.Cv, 04.20.Jb, 04.20.Dw}
\end{abstract}
%
\sloppy
\newcommand{\lfrac}[2]{{#1}/{#2}}
\newcommand{\sfrac}[2]{{\small \hbox{${\frac {#1} {#2}}$}}}
\newcommand{\ben}{\begin{eqnarray}}
\newcommand{\een}{\end{eqnarray}}
\newcommand{\la}{\label}
\maketitle
%
\section{Introduction}

This is the second of series articles in which we describe a new
geometrical models of general relativistic stars (GRS). One can
find the general scheme, notations, basic equations, additional
conditions and basic principles and properties of these models in
the first article of this series -- \cite{F04NMGS}. The
essentially new features of GRS in these new models are not based
on the critics or revision of very general relativity (GR). They
are a result of more deep understanding of its applications, and
on solution of some open problems in this theory, like the
physical justification of the choice of GR gauges. The preliminary
knowledge of the article \cite{F04NMGS} is highly recommended. It
is essential for the right understanding of the present one.

Here we consider the simplest specific example of application of
the general scheme, described in \cite{F04NMGS}: GRS, made of an
incompressible matter with equation of state (EOS):
\ben \varepsilon(p)=const=\varepsilon. \la{inc}\een
In Eq. (\ref{inc}) $\varepsilon$ is the energy density and $p$ is
the pressure of stelar matter. We are using units $c=G_N=1$.
Further on the letter $"C"$, as an index of different quantities,
denotes their values at the center of the star, the sign $"*"$ --
the values at the edge of the star.

Such simple model has a limited physical significance, because it
leads to an infinite speed of the sound in the fluid \cite{books}.
Nevertheless, its consideration is useful, because:

$\bullet$ It gives some general estimations for the properties of
the solutions of  extended Tolman-Openheimer-Volkov (ETOV)
equations with arbitrary EOS \cite{books}.

$\bullet$ This simple model can be used as a good physical
approximation for description of neutron stars with matter density
$\approx 2.85\times 10^{14}\,\,g/cm^3$ and $p_C\lesssim 5\times
10^{33}\,\,erg/cm^3$, because under these conditions the nuclear
matter behaves much like incompressible fluid \cite{books}.

$\bullet$ The EOS  (\ref{inc}) has the advantage that it yields an
analytically solvable model of stars. The simplest {\em
degenerate} set of solutions with luminosity variable $\rho_C=0$
was at first found in the Schwarzschild pioneer article
\cite{Schwarzschild}.

$\bullet$ It is a basic example of application of GR to the stelar
physics \cite{books}.

Because of the overestimation of the role of the luminosity
variable $\rho$, up to now the considerations of GRS structure
were based, as a rule, on the Hilbert gauge (HG): $\rho(r)\equiv
r$, in which the condition $\rho_C=0$ seems to be natural. Here
$r$ is the proper radial variable of the spherically symmetric
problem at hand \cite{F04NMGS}.

To the best of our knowledge, the solutions with $\rho_C>0$ have
not been studied and do not have a proper physical interpretation.
We intend to fill this gap in the present article. The general
solution of this problem, described here, turns to be much more
complicated and more interesting then the degenerate one,
considered by Schwarzschild in \cite{Schwarzschild}.

The new models of GRS recover an essentially new and unexpected
relativistic physics. In particular, in these models GRS with {\em
arbitrary large} mass $m_*$ are allowed. These are to have
arbitrary small geometrical radius $R_*$ and arbitrary small
luminosity. We refer to such amazing relativistic objects as {\em
heavy black dwarfs} (HBD).

In the present article we give a mathematical proves of the
existence of HBD in the specific case of incompressible matter.
Taking into account that the EOS has a quite week influence on the
proper radial gauge $\rho(r)$ \cite{F04NMGS}, one may expect this
result to be general.

Pure GR reasons, which are independent of EOS, can not yield
restrictions on the maximal mass of stars. Constraints of that
kind may arise only due to quantum statistics via the EOS, and in
equal footing both in Newton gravity \cite{books, Chandra} and in
GR. These theoretical conclusions may give a more precise physical
understanding of the real phenomena in stelar physics and need a
further study.

The new family of GRS is two parameter one. As a free parameters
one can consider, for example, the mass $m_*$ and the radius $R_*$
\cite{F04NMGS}. Between these quantities, in general, there exist
no stiff functional relation, as in the widely accepted GR models
with extra condition $\rho_C=0$ \cite{books, WD}. Actually, just
this assumption is responsible for existence of the well known
mass-radii relations \cite{F04NMGS}. In a framework of a given
theory of gravity (Newtonian, GR, etc) their specific form depends
on EOS, but their common origin is in the condition $\rho_C=0$.

The very recent observational data (see J.~Madej et al., 2004 in
\cite{WD}) seem to confirm the existence of two-parameter family
of white dwarfs with free parameters $m_*$ and $R_*$ in proper
domain, thus rejecting the existence of stiff mass-radii relations
in Nature. Here we give a preliminary comparison of these data
with our new models of GRS, using the results, obtained for the
simple case of incompressible stars.

Making use of the simplest EOS  (\ref{inc}) we will be able to
illustrate in most transparent way the differences between our
approach to the GRS structure and the commonly accepted one with
extra condition $\rho_C=0$.

\section{General Solution of ETOV Equations for
            Incompressible Matter in Hilbert Gauge}

For EOS (\ref{inc}) the equation
\ben {{dm}\over{d\rho}}=4\pi\varepsilon\rho^2
> 0, \hskip 1.truecm\\
m(\rho_C,\rho_C)=0 ,\,\,\,\,m(\rho_*,\rho_C)=m_*
\nonumber\la{TOV1}\een
splits from the whole ETOV system of differential equations and
has a simple general solution:
\ben m(\rho,\rho_C)={4\over 3}
\pi\varepsilon\left(\rho^3-\rho_C^3\right). \la{solm}\een
The general solution of the equation
\ben {{dp}\over{d\rho}}=-{{(p+\varepsilon)(m+4\pi\rho^3 p)}
\over{\rho(\rho-2m)}} < 0, \\
p(\rho_C,\rho_C)=p_C ,\,\,\,\,p(\rho_*,\rho_C)=0\hskip .2truecm
\nonumber\la{TOV2} \een
can be written in the form
\ben p(\rho,\rho_C)+\varepsilon=
{{\sqrt{-g_{\rho\rho}(\rho,\rho_C)}\left(p_C+\varepsilon\right)}\over
{1+\left(p_C/\varepsilon+1\right)
\chi(\rho,\rho_C)}}.\la{solp}\een
The general solutions of the equations

\ben
{{dm_0}\over{d\rho}}=4\pi\varepsilon\rho^2\sqrt{-g_{\rho\rho}} >
0, \hskip .8truecm\\
m_0(\rho_C,\rho_C)=0 ,\,\,\,\,m_0(\rho_*,\rho_C)=m_{0*},
\nonumber\la{TOV3}\een
and
\ben {{d\varphi}\over{d\rho}}={{m+4\pi\rho^3 p}
\over{\rho(\rho-2m)}}>0,\hskip .5truecm\\
\varphi(\rho_C,\rho_C)\!=\!\varphi_C
,\,\varphi(\rho_*,\rho_C)\!=\!\varphi_* \nonumber \la{TOV4}\een

have the form
\ben m_0(\rho,\rho_C)=4\pi\varepsilon\int\limits^\rho_{\rho_C}
\rho^2\sqrt{-g_{\rho\rho}(\rho,\rho_C)}\, d\rho,\la{solm0}\een
\ben \varphi(\rho,\rho_C)=\ln\left({{1+\left(w_C+1\right)
\chi(\rho,\rho_C)}\over
{\sqrt{-g_{\rho\rho}(\rho,\rho_C)}}}\right) +\varphi_C.
\la{solphi}\een

In the above formulas \ben
\chi(\rho,\rho_C)=4\pi\varepsilon\int\limits_{\rho_C}^\rho
\rho\left(\sqrt{-g_{\rho\rho}(\rho,\rho_C)}\right)^3 d\rho.
\la{chi}\een
 The solution (\ref{solp}) can be rewritten in a more convenient
 form \ben\chi(\rho,\rho_C)={{\sqrt{-g_{\rho\rho}}}\over{w+1}}-
\left({{\sqrt{-g_{\rho\rho}}}\over{w+1}}\right)_C \la{solchi}\een
making use of the scale-invariant ratio
$w=p/\varepsilon$\cite{F04NMGS}.

Then the metric coefficients can be written in the form:
\ben g_{tt}(\rho,\rho_C)\!=\!e^{2\varphi(\rho,\rho_C)}
,\nonumber \\
g_{\rho\rho}(\rho,\rho_C)\!=\!{-1\over{1\!-\!2m(
\rho,\rho_C)/\rho}}. \la{sg}\een

\section{Scale Properties of the Problem and HG Scale Invariant
Quantities for Incompressible Stars}

The EOS  (\ref{inc}) is not scale-invariant. Since the scale
factor is $\lambda=const$, the scale transformations:
\ben \rho\!\to\!\lambda\rho,\,m\!\to\!\lambda
m,\,m_0\!\to\!\lambda m_0
,\,\varepsilon\!\to\!\lambda^{-2}\varepsilon,\,p\!\to\!\lambda^{-2}
p.\hskip .6truecm\la{lambda}\een
described in \cite{F04NMGS, developments} in more details, map the
solution of ETOV system with a fixed value of the constant density
$\varepsilon$ onto another solution of the same system with a
constant energy density $\varepsilon^\prime =
\lambda^{-2}\varepsilon$. Utilizing this property one can consider
only the case $\varepsilon=1$. Instead, we prefer to use a scale
invariant variables. In contrast to other authors
\cite{developments}, we avoid the {\em a prior} choice of such
variables and will find them during the very course of the
solution of ETOV system with EOS (\ref{inc}).

In the case of incompressible star one can use the representation
\ben g_{\rho\rho}(\rho,\rho_C)={3\over
{8\pi\varepsilon}}\rho^2/P_4(\rho,\rho_C)\la{gP4},\een where \ben
P_4(\rho,\rho_C):=\rho\left(\rho^3-{{3}\over{8\pi\varepsilon}}\rho-
\rho_C^3\right)= \prod_{i=1}^4(\rho-\rho_i)\hskip
.7truecm\la{P4rho}\een is a naturally normalized polynomial of
fourth degree, which roots are $\rho_{1,2,3,4}$. It is easy to
find these roots in the form \ben
\rho_1={{1}\over{\sqrt{2\pi\varepsilon\left(1+\sigma^2/3\right)}}}>\rho_2=0>
\hskip 2.9truecm\nonumber\\
\rho_3={{-{1\over 2 }(1-\sigma)}
\over{\sqrt{2\pi\varepsilon\left(1+\sigma^2/3\right)}}}>
\rho_4={{-{1\over 2 }(1+\sigma)}
\over{\sqrt{2\pi\varepsilon\left(1+\sigma^2/3\right)}}},\hskip
.8truecm \la{rootsrho}\een where the $\lambda$-invariant parameter
$\sigma$ is defined via the substitution
$\rho_C={{1}\over{\sqrt{2\pi\varepsilon
\left(1+\sigma^2/3\right)}}}\left({{1-\sigma^2}\over{4}}\right)^{1/3}$
\footnote{Note that in the units $c=G_N=1$ the energy density
$\varepsilon$ has a dimension of {\em length}${}^{-2}$.}.

Now it becomes clear that instead of the luminosity variable
$\rho$ it is better to use the $\lambda$-invariant one, $\xi$:
\ben \xi:= \sqrt{2\pi\varepsilon \left(1+\sigma^2/3\right)}
\,\rho=\sqrt{{{8\pi\varepsilon}\over 3}
\Big(1-\xi_C^3\Big)}\,\rho,\la{xi}\een
where $\xi_C\in (0,1)$ is the only positive root of the cubic
equation $\xi_C^3+{{3}\over{8\pi\varepsilon\rho_C^2}}\xi_C^2-1=0$
($\Rightarrow\sigma=\sqrt{1-4\xi_C^3}$).

As a result we obtain
\ben g_{\rho\rho}(\xi,\xi_C)=\left(1-\xi_C^3\right)
{{\xi^2}\over{P_4(\xi,\xi_C)}}.\la{gxi}\een

The naturally re-normalized $\lambda$-invariant polynomial
\ben
P_4(\xi,\xi_C)\!=\!\prod_{i=1}^4\left(\xi\!-\!\xi_i\right)\!=\!
\xi(\xi-1)\Big(\xi^2\!+\!\xi\!+\!\xi_C^3\Big)\hskip
.8truecm\la{P4xi}\een
\begin{figure}[htbp] \vspace{5.truecm}
\includegraphics{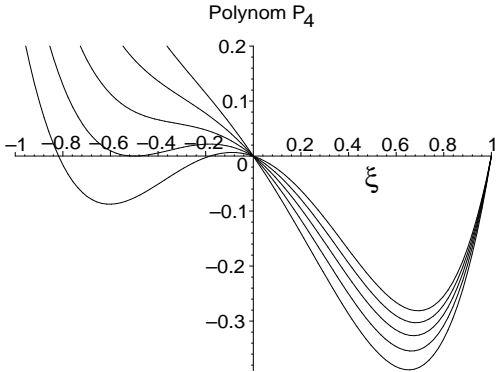} \caption{\hskip 0.2truecm The polynomial
$P_4(\xi,\xi_C)$ for different values of the parameter $\xi_C$.
    \hskip 1truecm}
    \label{Fig1}
\end{figure}
has the following re-scaled roots:
\ben \xi_1\!=\!1>\xi_2\!=\!0>\hskip
3.7truecm\nonumber \\
\xi_3\!=\!-{{1\!-\!\sqrt{1\!-\!4\xi_C^3}}\over 2}
>\xi_4\!=\!-{{1\!+\!\sqrt{1\!-\!4\xi_C^3}}\over
2}\nonumber\\ \hbox{-- if}\,\,\, \xi_C^3 \in [0,{1/ 4}],\hskip
3.5truecm\la{rootxiRe}\een
and
\ben \xi_1\!=\!1>\xi_2\!=\!0>\hskip
4.4truecm\nonumber \\
\xi_3\!=\!-{{1\!-\!i\sqrt{|1\!-\!4\xi_C^3}|}\over 2}
>\xi_4\!=\!-{{1\!+\!i\sqrt{|1\!-\!4\xi_C^3}|}\over
2}\nonumber\\ \hbox{-- if}\,\,\, \xi_C^3 \in [{1/ 4},1].\hskip
4.15truecm\la{rootxiIm}\een

There exist two degenerate cases:

a) Schwarzschild degenerate case: $\xi_C\!=\!0$\,$\Rightarrow$\,
$\xi_2\!=\!\xi_3\!=\!0$ is a double root and
$P_4(\xi,0)=-\xi^2(1-\xi^2)$;

b) A new degenerate case: $\xi_C\!=\!\sqrt[3]{1/4}$ $\Rightarrow $
$\xi_3\!=\!\xi_4\!=-{1\over2}$ is a double root and
$P_4(\xi,0)=-\xi(1-\xi)(\xi-1/2)^2$.

In both cases the corresponding elliptic integrals, needed for an
explicit solution of the problem, are reduced to elementary
functions -- see Appendix A.

For the $\lambda$-invariant quantities
\ben \mu(\xi,\xi_C)=\sqrt{{{8\pi\varepsilon}\over 3}
\Big(1-\xi_C^3\Big)}\,m(\rho,\rho_C),\nonumber\\
\mu_0(\xi,\xi_C)=\sqrt{{{8\pi\varepsilon}\over 3}
\Big(1-\xi_C^3\Big)}\,m_0(\rho,\rho_C)\,\la{mumu0}\een
from Eq.(\ref{solm}), (\ref{solm0}) we obtain (see Appendix A)\ben
\mu(\xi,\xi_C)={1\over 2}{{\xi^3-\xi_C^3}\over{1-\xi_C^3}},
\nonumber\\ \mu_0(\xi,\xi_C)={3\over
2}{{1}\over{\sqrt{1-\xi_C^3}}} \int\limits^\xi_{\xi_C}{{\xi^3 d\xi
}\over{\sqrt{-P_4(\xi,\xi_C)}}} =\nonumber\\{3\over
2}{{1}\over{\sqrt{1-\xi_C^3}}}
\Big(J_{3|1}(\xi,\xi_C)-J_{3|1}(\xi_C,\xi_C)\Big).\hskip .1truecm
\la{solmumu0}\een

\begin{figure}[htbp] \vspace{7.truecm} \includegraphics{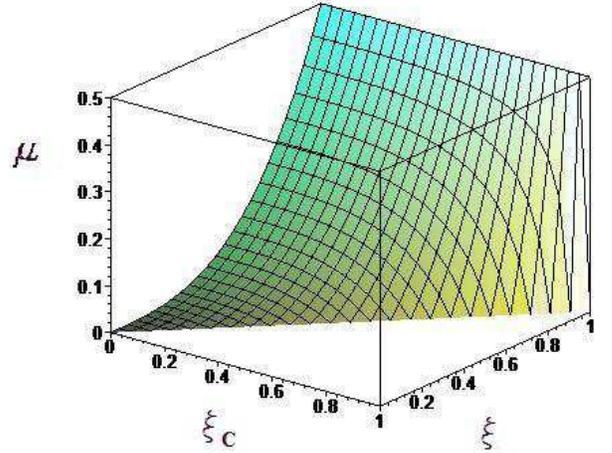} \caption{\hskip
0.2truecm The function $\mu(\xi;\xi_C)\in[0,1/2]$ and the lines
$\xi=const$ and $\xi_C=const$.
    \hskip 1truecm}
    \label{Fig2}
\end{figure}

\begin{figure}[htbp] \vspace{8.truecm} \includegraphics{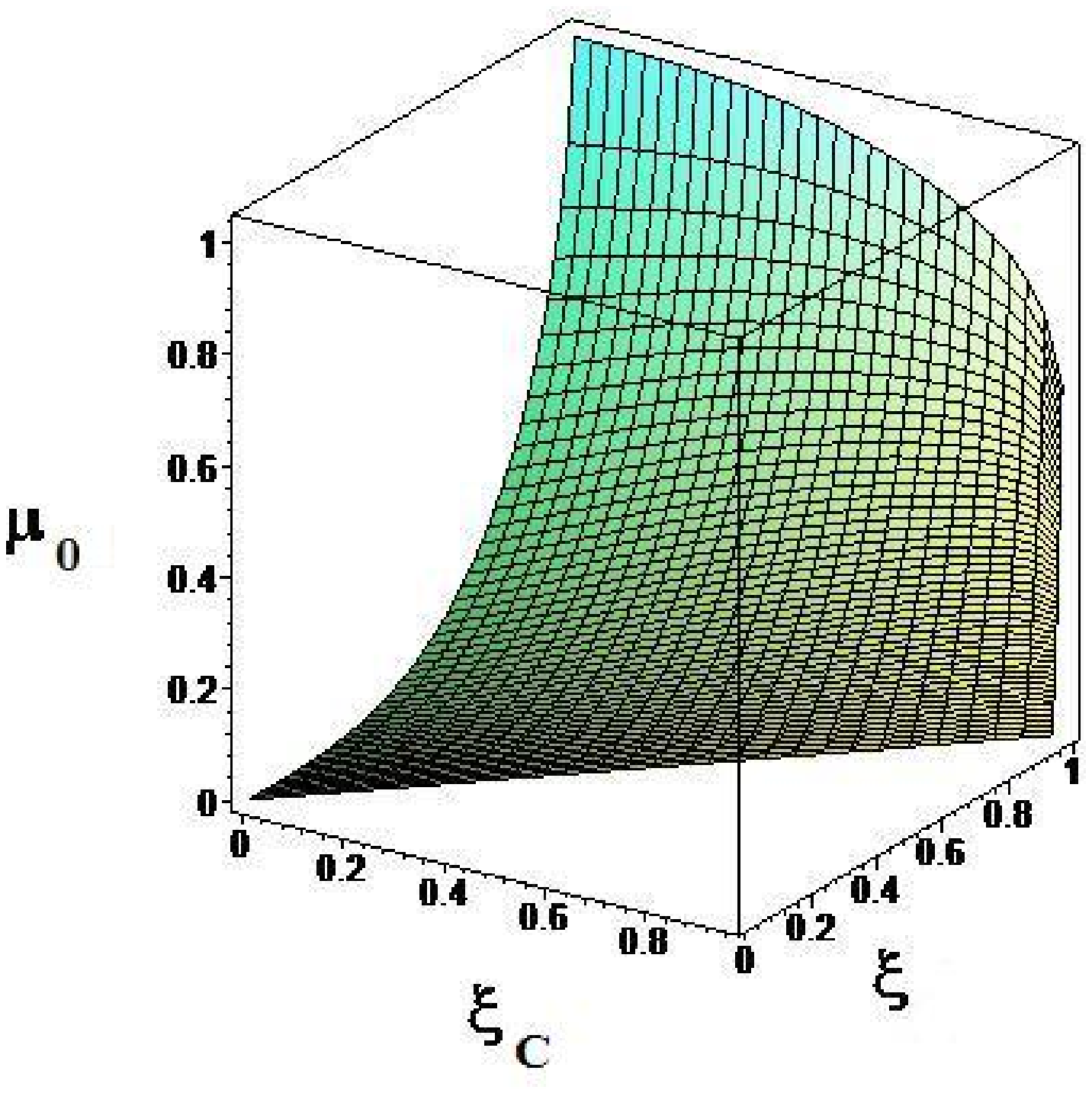} \caption{\hskip
0.2truecm The function $\mu_{0}(\xi;\xi_C)$ and the lines
$\xi=const$ and $\xi_C=const$.
    \hskip 1truecm}
    \label{Fig2b}
\end{figure}

Instead of the local binding energy $\Delta
m(\rho,\rho_C):=m_0-m$, which is not $\lambda$ -invariant, one can
consider the ratio
\ben\varrho(\rho,\rho_C):=m/m_0=\mu/\mu_0=\varrho(\xi,\xi_C)\in
(0,1).\hskip.5truecm\la{varrho}\een

It measures in a $\lambda$ -invariant way the local mass defect of
the star mater, i.e. the mass defect in the sphere with luminosity
radius $\rho$ and center $C$.

Another important $\lambda$-invariant local (in the above sense)
quantity is
$f(\rho,\rho_C)=\varrho(\rho,\rho_C)^2-g_{tt}(\rho,\rho_C)+1$. In
the case at hand it has the form
\ben f(\rho,\rho_C)=\left({{m}\over{m_0}}\right)^2+{{2
m}\over{\rho}}=\varrho^2+\varsigma^2,\la{Frho}\een
where $\varsigma^2\!=\!{{2 m}\over{\rho}}\geq 0$ is the local
compactness of the star. For incompressible stars
$\varsigma^2={{\xi-\xi_C^3}\over{\xi(1-\xi_C^3)}}$.

Using the results (\ref{J31}) and (\ref{J43}) of Appendix A, after
some algebra one obtains the expressions:
\ben
\varrho(\xi,\xi_C)^{-1}=3\,{{\sqrt{1-\xi_C^3}}\over{\xi^3-\xi_C^3}}
\int\limits_{\xi_C}^\xi {{\xi^3 d\xi
}\over{\sqrt{-P_4(\xi,\xi_C)}}} =\nonumber\\
{3}{{\sqrt{1-\xi_C^3}}\over{\xi^3-\xi_C^3}}
\Big(J_{3|1}(\xi,\xi_C)-J_{3|1}(\xi_C,\xi_C)\Big) \hskip .5truecm
\la{varrhoxi}\een
and
\ben \chi(\xi,\xi_C)= {3\over
2}\sqrt{1-\xi_C^3}\int\limits_{\xi_C}^\xi {{\xi^4
d\xi}\over{\left(\sqrt{-P_4(\xi,\xi_C)}\right)^3}}=\nonumber\\
{3\over
2}\sqrt{1-\xi_C^3}\Big(J_{4|3}(\xi,\xi_C)-J_{4|3}(\xi_C,\xi_C)\Big)\hskip
1truecm \la{chixi}\een
for the basic $\lambda$-invariant local quantities $\varrho$ and
$\chi$.
\begin{figure}[here] \vspace{7.5truecm} \includegraphics{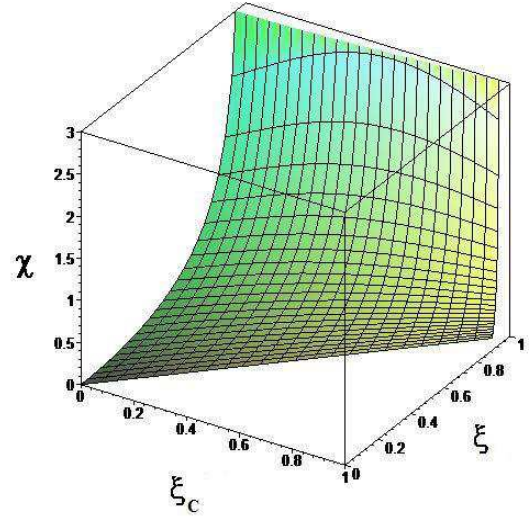} \caption{\hskip
0.2truecm The function $\chi(\xi;\xi_C)$ and the lines $\xi=const$
and $\xi_C=const$. In contrast to $m_{0}(\xi;\xi_C)$, the function
$\chi(\xi;\xi_C)$ is not bounded from above.
    \hskip 1truecm}
    \label{Fig2c}
\end{figure}

As one sees, the technical problem of finding the general solution
of ETOV system for incompressible matter in the most general case
is reduced to the calculation of the elliptic integrals in Eqs.
(\ref{mumu0}), (\ref{varrhoxi}) and (\ref{chixi}). This
calculation is described in the Appendix A.

\section{Radius of Incompressible Stars}

Now we are able to proof the existence of a finite coordinate
radius $r_*<\infty$ of incompressible stars. It corresponds to
some finite value of the luminosity variable $\xi_*<\infty$ and to
some finite {\em geometrical} radius $R_*={\cal
R}_*/\sqrt{8\pi\varepsilon\left(1-\xi_C^3\right)/3}<\infty$. (Here
${\cal R}_*$ is the dimensionless $\lambda$-invariant geometrical
radius of the star.)

By definition, in $\lambda$-invariant terms the edge of the star
is defined as a point $\xi_*$ at which \ben
p(\xi_*,\xi_C)=0.\la{xi*}\een

{\bf\em Proposition 1:} {\em For all solutions of ETOV equations}
(\ref{TOV1})-(\ref{TOV3}) {\em with EOS} (\ref{inc}) {\em and
arbitrary $\xi_C\geq 0$ there exist a unique solution
$\xi_*\in\big(\xi_C,\xi_*^{crit}(\xi_C)\big)$ of the Eq.}
(\ref{xi*}). {\em Here the value $\xi_*^{crit}(\xi_C)\in(\xi_C,1)$
corresponds to the nonphysical limiting solution with infinite
value of the pressure at the stelar center: $p_C^{crit}=\infty$.}

{\bf\em Proof:}

1) Existence of unique solution of Eq. (\ref{xi*}):

From Eq. (\ref{gxi}) and (\ref{chixi}) one easily obtains in the
limit $\xi\to 1-0$: \ben
\sqrt{-g_{\rho\rho}(\xi,\xi_C)}\,\propto\,
\sqrt{{{1-\xi_C^3}\over{2+\xi_C^3}}{{1}\over{1-\xi}}}\to\infty, \nonumber\\
\chi(\xi,\xi_C)\,\propto\,
3\,\sqrt{{{1-\xi_C^3}\over{\left(2+\xi_C^3\right)^3}}{{1}\over{1-\xi}}}\to\infty.
\la{gchilim}\een Hence \ben \lim_{\xi\to 1-0}
{{\sqrt{-g_{\rho\rho}(\xi,\xi_C)}}\over{\chi(\xi,\xi_C)}}={{2+\xi_C^3}\over{3}}\in
(2/3,1). \la{lim_sgf}\een  As a result \ben \lim_{\xi\to 1-0}
w(\xi,\xi_C)=-{{1-\xi_C^3}\over{3}}<0, \la{lim_w}\een i.e. the
continuous, strictly monotonic function $w(\xi,\xi_C)$ decreases
on the interval $[\xi_C,1]$ from some value $w_C=w(\xi_C,\xi_C)>0$
to the value $w(1,\xi_C)=-{{1-\xi_C^3}\over{3}}<0$ (i.e.,
$w(\xi,\xi_C)\!\searrow\![w_C,\!-{(1\!-\!\xi_C^3)/{3}}]$ for
$\xi\in[\xi_C,1)$). Then there exist a unique value $\xi_*\in
(\xi_C,1)$, such that $w(\xi_*,\xi_C)=0$. This value $\xi_*$
defines the radius of the star.

2) Limitations on the $\lambda$-invariant radius of star $\xi_*$:

The Eq. (\ref{solchi}) permits us to rewrite the definition of the
stelar radius  (\ref{xi*}) in the following explicit form:
\ben w_C\!=\!{{1}\over{\sqrt{-g_{\rho\rho}(\xi_*,\xi_C)}
\!-\!\chi(\xi_*,\xi_C)}}\!-\!1\!=:w_C(\xi_*,\xi_C)\la{wC}.\hskip
.7truecm\een
\begin{figure}[htbp] \vspace{7.5truecm} \includegraphics{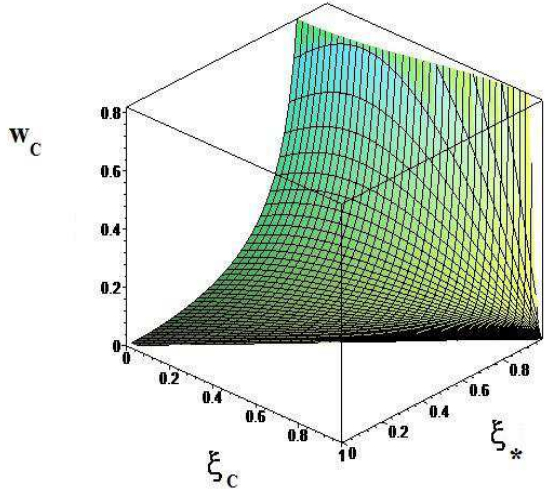} \caption{\hskip
0.2truecm The function $w_C(\xi_*;\xi_C)$ and the lines
$\xi_*=const$ and $\xi_C=const$. Note that
$w_C(\xi*;\xi_C)=\infty$ on some special line
$\xi_*=\xi_*^{crit}(\xi_C)$ which will be studied in details in
Subsection E of the present Section.
    \hskip 1truecm}
    \label{Fig2d}
\end{figure}

It demonstrates the relation between the value $w_C$ at the center
$C$ of the star and its radius $\xi_*$. This relation is the main
difference between relativistic theory of stars and Newtonian one.

Using the following properties of the functions
$\sqrt{-g_{\rho\rho}(\xi_*,\xi_C)}$ and $\chi(\xi_*,\xi_C)$

a) $-g_{\rho\rho}(\xi_C,\xi_C)=1$;

b) $\sqrt{-g_{\rho\rho}(\xi_*,\xi_C)} \nearrow [1,\infty)$,
$\chi(\xi_*,\xi_C)\nearrow [0,\infty)$ -- for $\xi\in[\xi_C,1)$;

c) for $0<\xi-\xi_C\ll 1$: $\sqrt{-g_{\rho\rho}(\xi_*,\xi_C)}
\!-\!\chi(\xi_*,\xi_C)>0$;

d) for $\xi\to 1$: $ \sqrt{-g_{\rho\rho}(\xi_*,\xi_C)} /
\chi(\xi_*,\xi_C) \propto (2+\xi_C^3)/3 \in (2/3,1)$; , one easily
obtains that:

i) $w_C(\xi_C,\xi_C)=0$;

ii) $w_C(1,\xi_C)=-1$;

ii) There exist a unique point $\xi_*^{crit}\in (\xi_C,1)$, such
that
$\sqrt{-g_{\rho\rho}(\xi_*^{crit},\xi_C)}=\chi(\xi_*^{crit},\xi_C)
\,\,\Rightarrow\,\,$ $w_C(\xi_*^{crit}-0,\xi_C)=+\infty$ and
$w_C(\xi_*^{crit}+0,\xi_C)=-\infty$.

This means that there exist a limiting nonphysical solution -- an
incompressible star with a critical radius $\xi_*^{crit}\in
(\xi_C,1)$, which corresponds to an infinite pressure
$p_C^{crit}=\infty$ at the center $C$. Hence, the
$\lambda$-invariant radius of the incompressible stars is
constraint in the interval $(\xi_C,\xi_*^{crit})$. It is obvious
that $\xi_*^{crit}$ is a function of $\xi_C$. At the end of this
Section we shall obtain the equation for determining of the
function $\xi_*^{crit}(\xi_C)$ and study its solution in details.

This completes the proof of our Proposition.

The luminosity radius $\rho_*$ of the star can be find in the form
$\rho_*={{\xi_*}\over{\xi_C}}\rho_C\geq\rho_C$ and is not bounded
from above if $\varepsilon\to0$, or/and for $\xi_C\to 1$.

Having in our disposal the $\lambda$-invariant radius $\xi_*$ of
the star, we are able to introduce another basic
$\lambda$-invariant characteristics:

i) The total Keplerian mass $\mu_*=\mu(\xi_*,\xi_C)\in (0,1/2)$ --
for $0\leq \xi_C\leq\xi_*< 1$, see Fig~\ref{Fig2}.

ii) The total proper mass $\mu_{0*}=\mu_0(\xi_*,\xi_C)>0$.

iii) The total mass defect ratio
$\varrho_*=\varrho(\xi_*,\xi_C)>0$.

iv) The total compactness
$\varsigma_*=\varsigma(\xi_*,\xi_C)\!=\!{{2m_*}/{\rho_*}}\!=
\!{{\xi_*^3-\xi_C^3}\over{\xi_*(1-\xi_C^3)}}\!\in\!(0,1)$, etc.

v) In addition we obtain the relations
\ben
\varphi_*=\varphi(\xi_*,\xi_C)=\ln\left(w_C+1\right)+\varphi_C\la{varphi*}\een
and (see Appendix A):
\ben {\cal R}_*\!=\!{\cal R}_*(\xi_*,\xi_C)\!=\!
\sqrt{1\!-\!\xi_C^3}\int_{\xi_C}^{\xi_*}\!\!{{\xi d\xi
}\over{\sqrt{-P_4(\xi,\xi_C)}}}=\nonumber\\
\sqrt{1\!-\!\xi_C^3}\Big(J_{1|1}(\xi,\xi_C)-J_{1|1}(\xi_C,\xi_C)\Big).
\hskip .5truecm\la{R*}\een It is easy to see that ${\cal
R}_*(\xi_*,\xi_C)\to\infty$ when $\xi_C=0$ and $\xi_*\to 1-0$.

\begin{figure}[htbp] \vspace{8.truecm} \includegraphics{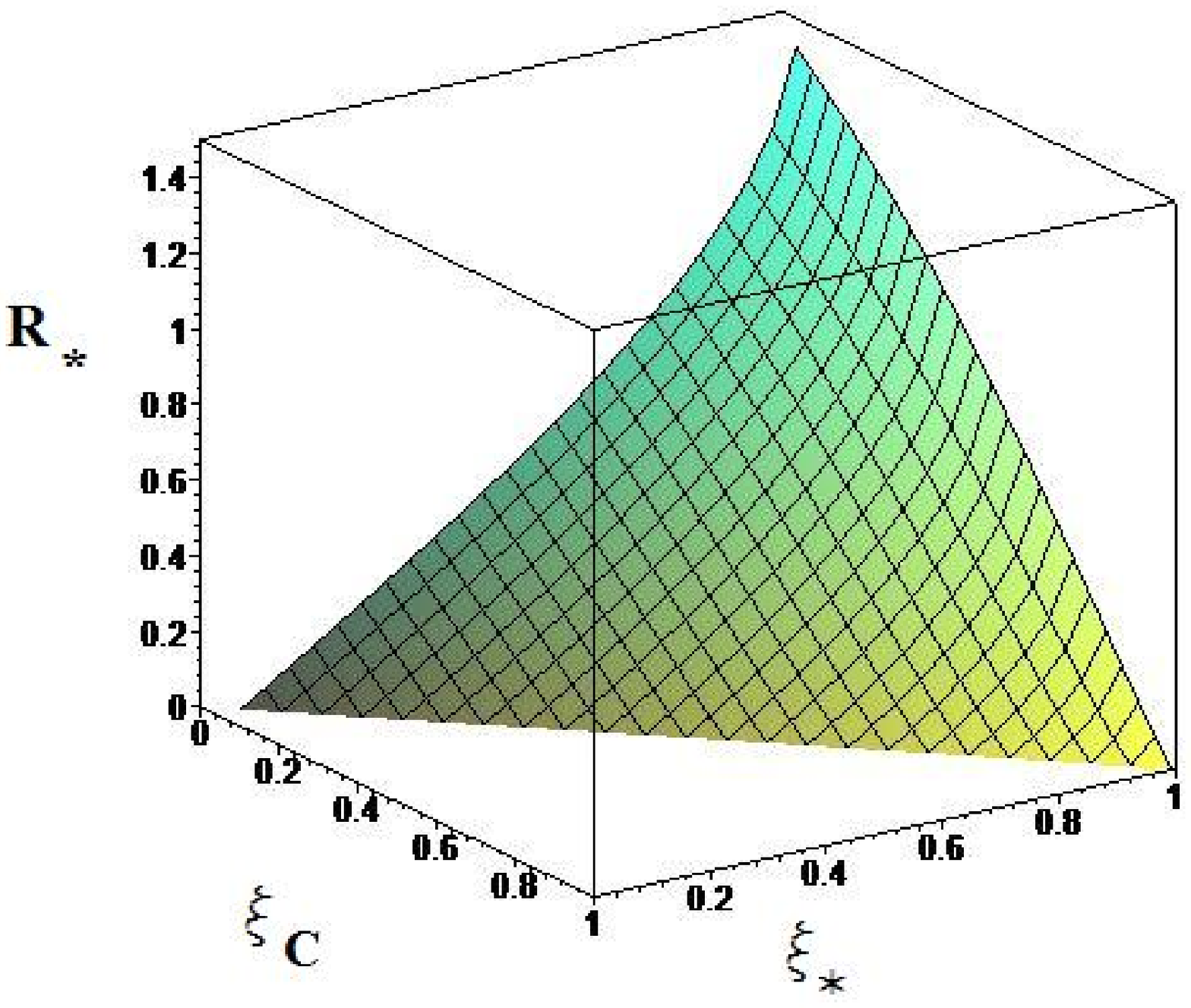} \caption{\hskip
0.2truecm A part of the function ${\cal R}_*(\xi_*;\xi_C)$ and the
lines $\xi_*=const$ and $\xi_C=const$.
    \hskip 1truecm}
    \label{Fig3}
\end{figure}

As seen in Fig. \ref{Fig2} the function $\mu_*(\xi_*,\xi_C)$ has
values in the interval $[0,1/2]$. At the same time we obtain a new

{\bf\em Proposition 2:} {\em In the specific limit: $\xi_*\to 1$,
$\xi_C\to 1$, $0\leq\xi_C\leq\xi_*\leq 1$ we have  \ben
\lim_{\underset{\xi_C\leq\xi_*\leq 1}{\xi_C\to 1}}
\mu_*(\xi_*,\xi_C)=:\mu_*(1\!-\!0,1\!-\!0)\in [0,1/2],\hskip
,5truecm \la{mu_limit}\een i.e., the limit
$\mu_*(1\!-\!0,1\!-\!0)$ is  bounded, but not definite and can
have any value in the interval $[0,1/2]$.}

The analytical proof becomes obvious from the representation
$\mu_*(\xi_*,\xi_C)={1\over 2 }{\over{1-\xi_c}}
{{\xi_*^2+\xi_*\xi_C+\xi_C^2}\over{1+\xi_C+\xi_C^2}}$ and
consideration of the limit $\xi_C\to 1$ of $\mu_*(\xi_*,\xi_C)$ on
the curves $\xi_*=\xi_C+k(1-\xi_C)$ with an arbitrary $k\in[0,1]$.

Note that for any $k\in[0,1]$ and $n>1,p>1$:

i) On the curves $\xi_*=\xi_C+k(1-\xi_C^p)^n$ the limit
(\ref{mu_limit}) has an universal value $0$.

ii) On the curves $\xi_*=1-k(1-\xi_C^p)^n$ the limit
(\ref{mu_limit}) has an universal value $1/2$.

Hence, the typical behavior of the limit (\ref{mu_limit}) is given
by the last two cases: i) and ii). The case of linear dependence
between $\xi_*$ and $\xi_C$ is an exceptional one.

The simple, but important property of the function
$\mu_*(\xi_*,\xi_C)$, described in Proposition 2,  will influence
further results in the theory of incompressible relativistic
stars, when the same limit of other quantities will emerge.

\section{The Basic $Rm$ Mapping for Incompressible Stars}

The basic mapping
\ben \{{\cal R}_*,\mu_*\} \xymatrix@1{ {} \ar[r]^{Rm} & {}}
\{\xi_*({\cal R}_*,\mu_*),\xi_C({\cal R}_*,\mu_*)\}. \la{map}\een
was defined in general case in \cite{F04NMGS}. Here we study this
mapping in the specific case of incompressible GRS in
scale-invariant variables:

Substituting in the formula (\ref{R*}) $\xi_*=\xi_*(\mu_*,\xi_C)$,
or $\xi_C=\xi_C(\mu_*,\xi_*)$, obtained from the first of the
relations (\ref{solmumu0}) in the form
\begin{subequations}\label{muxiCxiS:ab}
\ben
\xi_*(\mu_*,\xi_C)=\sqrt[3]{2\mu_*+(1-2\mu_*)\xi_C^3},\la{muxiCxiS:a}\\
\xi_C(\mu_*,\xi_*)=\sqrt[3]{(\xi_*^3-2\mu_*)/(1-2\mu_*)},\la{muxiCxiS:b}
\een
\end{subequations}
one obtains the functions ${\cal R}_*={\cal R}_*(\mu_*,\xi_C)$ and
${\cal R}_*={\cal R}_*(\mu_*,\xi_*)$ .

\begin{figure}[htbp] \vspace{5.5truecm} \includegraphics{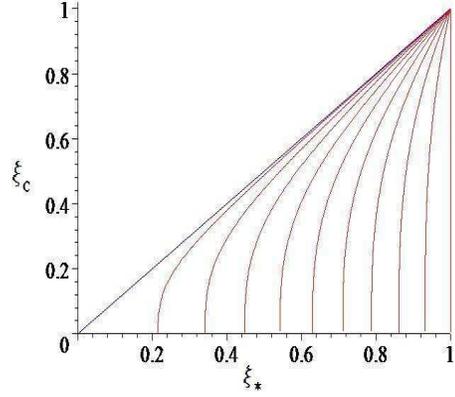} \caption{\hskip
0.2truecm The function $\xi_C=\xi_C(\mu_*,\xi_*)$
(\ref{muxiCxiS:b}) for different fixed values of parameter
$\mu_*\in (0,1/2)$.
    \hskip 1truecm}
    \label{Fig4}
\end{figure}

\begin{figure}[htbp] \vspace{7.truecm} \includegraphics{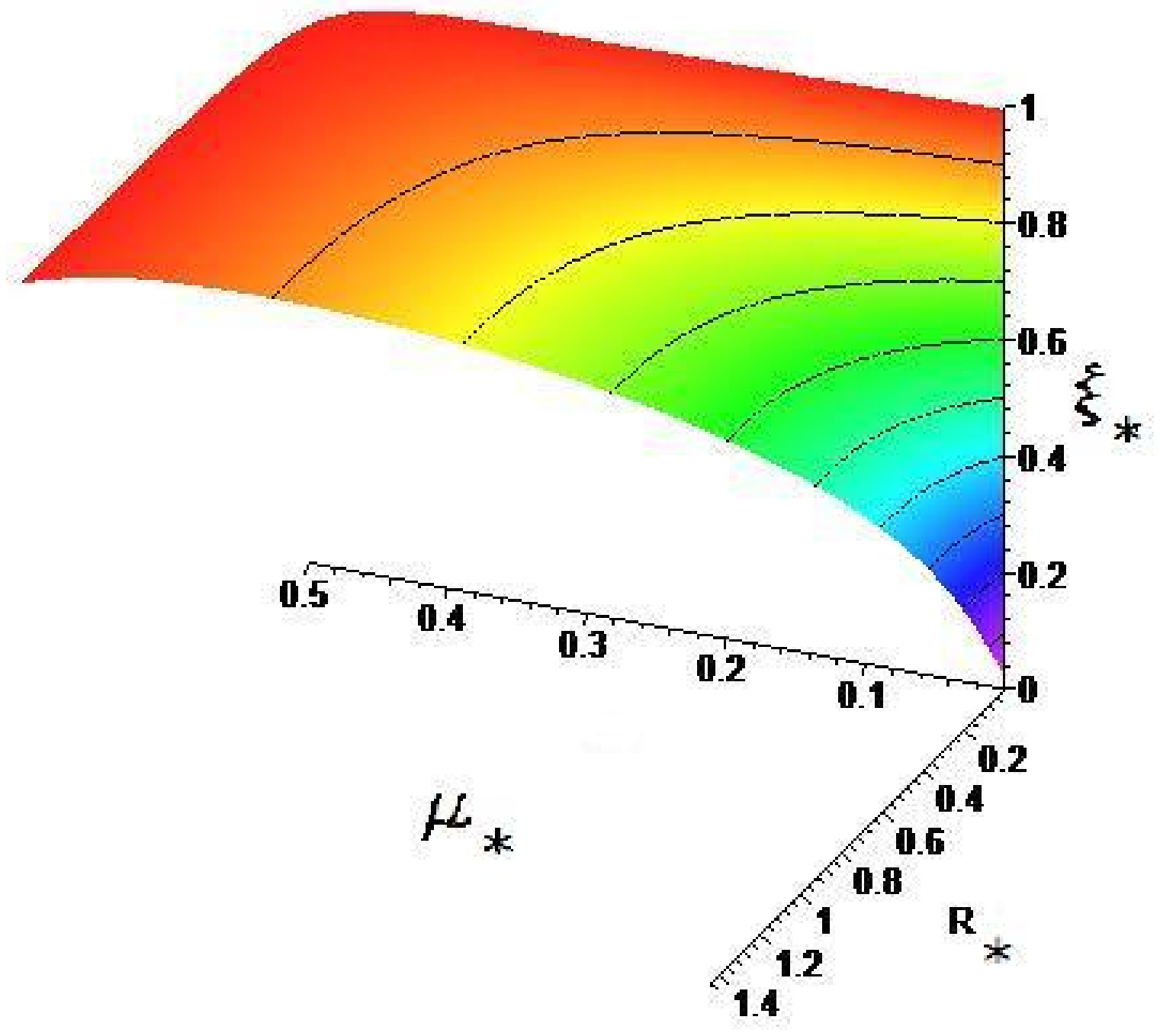} \caption{\hskip
0.2truecm The function $\xi_*(\mu_*;{\cal R}_*)\in[0,1]$. The
behavior of this function in the limit $\mu_*\to 0, {\cal
R}_*\to0$ is similar to the one, described in Proposition 2 for
the function $\mu_*(\xi_*,\xi_C)$.
    \hskip 1truecm}
    \label{Fig5}
\end{figure}

\begin{figure}[htbp] \vspace{7.truecm} \includegraphics{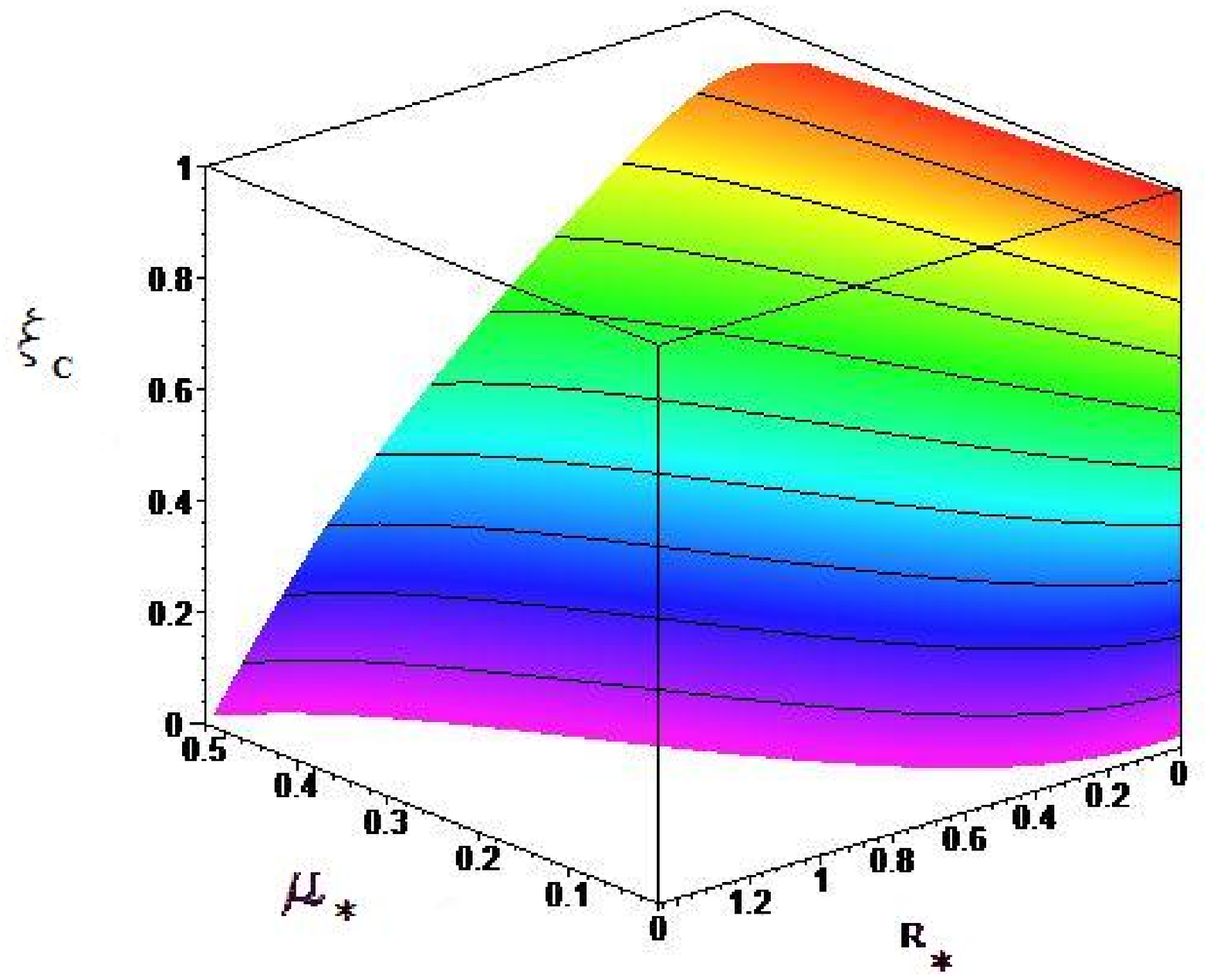} \caption{\hskip
0.2truecm The function $\xi_C(\mu_*;{\cal R}_*)\in[0,1]$. The
behavior of this function in the limit $\mu_*\to 0, {\cal
R}_*\to0$ is similar to the one, described in Proposition 2 for
the function $\mu_*(\xi_*,\xi_C)$.
    \hskip 1truecm}
    \label{Fig6}
\end{figure}

The corresponding inverse functions $\xi_*=\xi_*(\mu_*,{\cal
R}_*)$ and $\xi_C=\xi_C(\mu_*,{\cal R}_*)$ define the basic
mapping $Rm$ (\ref{map}) in $\lambda$-invariant terms. This
mapping is illustrated in figures \ref{Fig5} and \ref{Fig6}.

As a byproduct from Eq. (\ref{muxiCxiS:b}), $2\mu_*\leq 1$ and
$\xi_C\geq 0$ we obtain a new inequality $\xi_*^3\geq 2\mu_*$.
Then, since $0<\xi_*<1$, we have
\ben 0<{{2\mu_*}\over{\xi_*}}<{{2\mu_*}\over{\xi_*^3}}\leq 1.
\la{inequalities_mu*xi*}\een

For incompressible stars the relations, which determine the domain
${\mathbb D}^{(2)}_{\xi_*\xi_C}$ \cite{F04NMGS}, acquire the
following specific form (For the used notations see Appendix A.):
\begin{subequations}\label{Dxixi:xyz}
\ben
\renewcommand{\theequation}{\theparentequation\alpha{equation}}
 0\,\,\leq\,\,\xi_C\,\,\leq\,\,\xi_*\,\,<\,\,1,\hskip 4.5truecm \la{Dxixi:x}\\
-\xi_*(1+\xi_*)\,\,\leq\,\, 0\leq\,\,\xi_C^3,\hskip 4.5truecm   \la{Dxixi:y}\\
\sqrt{{{\xi_*}\over{(1-\xi_*)(\xi_C^3+\xi_*^2+\xi_*)}}}-
{{1}\over{\sqrt{1-\xi_C^3}}}\,\,\,<\hskip 1.8truecm\nonumber\\
{3\over 2}\Big(J_{4|3}\left(\xi_*,\xi_C\right))-
J_{4|3}\left(\xi_C,\xi_C\right)\Big)\!<\!\hskip 1.8truecm \nonumber\\
\sqrt{{{\xi_*}\over{(1\!-\!\xi_*)(\xi_C^3+\xi_*^2+\xi_*)}}}\hskip
2.5truecm.\la{Dxixi:z} \een
\end{subequations}

As a result of the first of the relations (\ref{Dxixi:xyz}) the
condition (\ref{Dxixi:y}) is fulfilled.

\begin{figure}[htbp] \vspace{7.5truecm}
\includegraphics{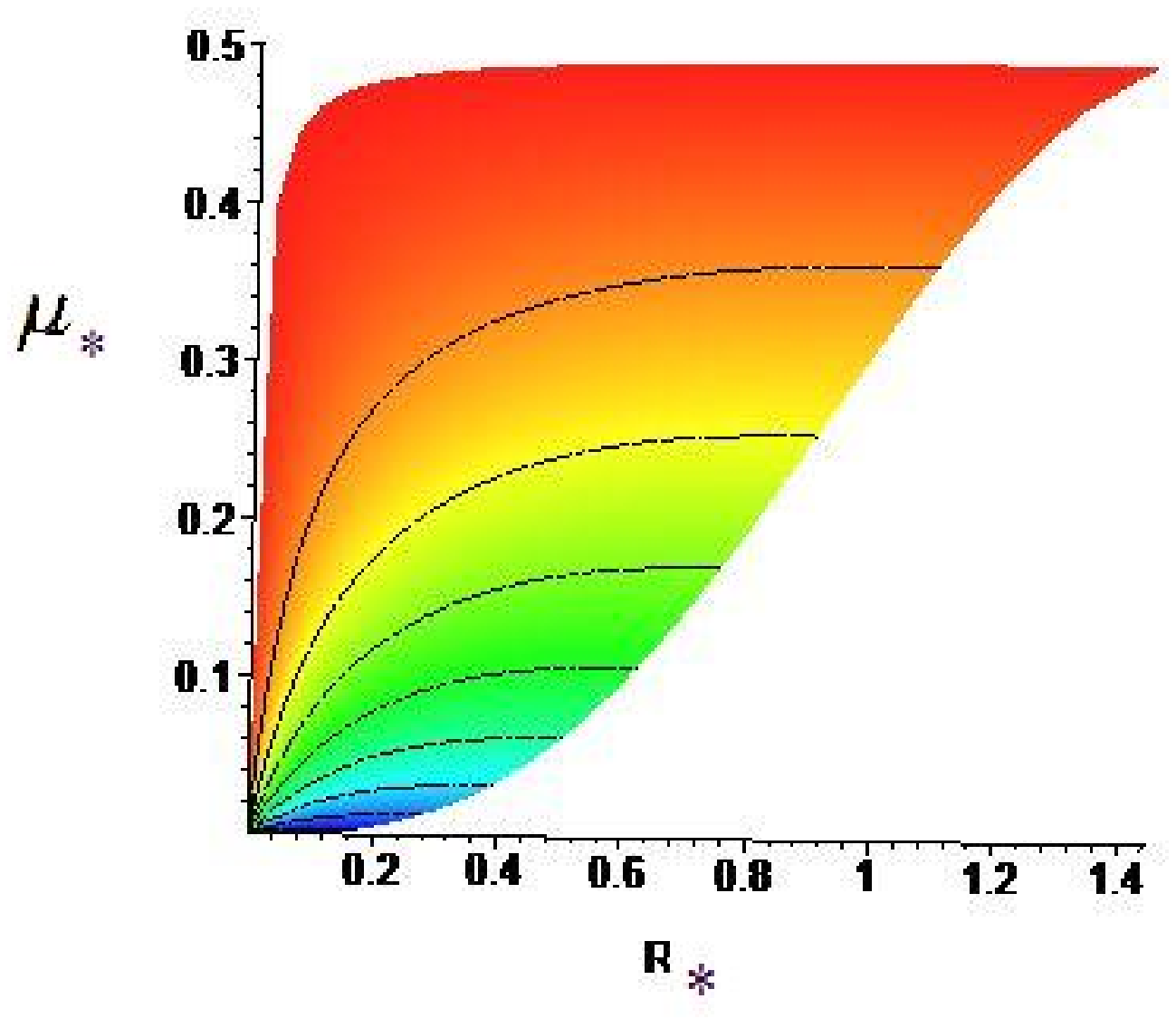} \caption{\hskip 0.2truecm The domain ${\mathbb
D}^{(2)}_{\mu_*,R_*}$ for the function $\xi_C(\mu_*;{\cal R}_*)$.
    \hskip 1truecm}
    \label{Fig7}
\end{figure}

\begin{figure}[htbp]\vspace{7.5truecm}
\includegraphics{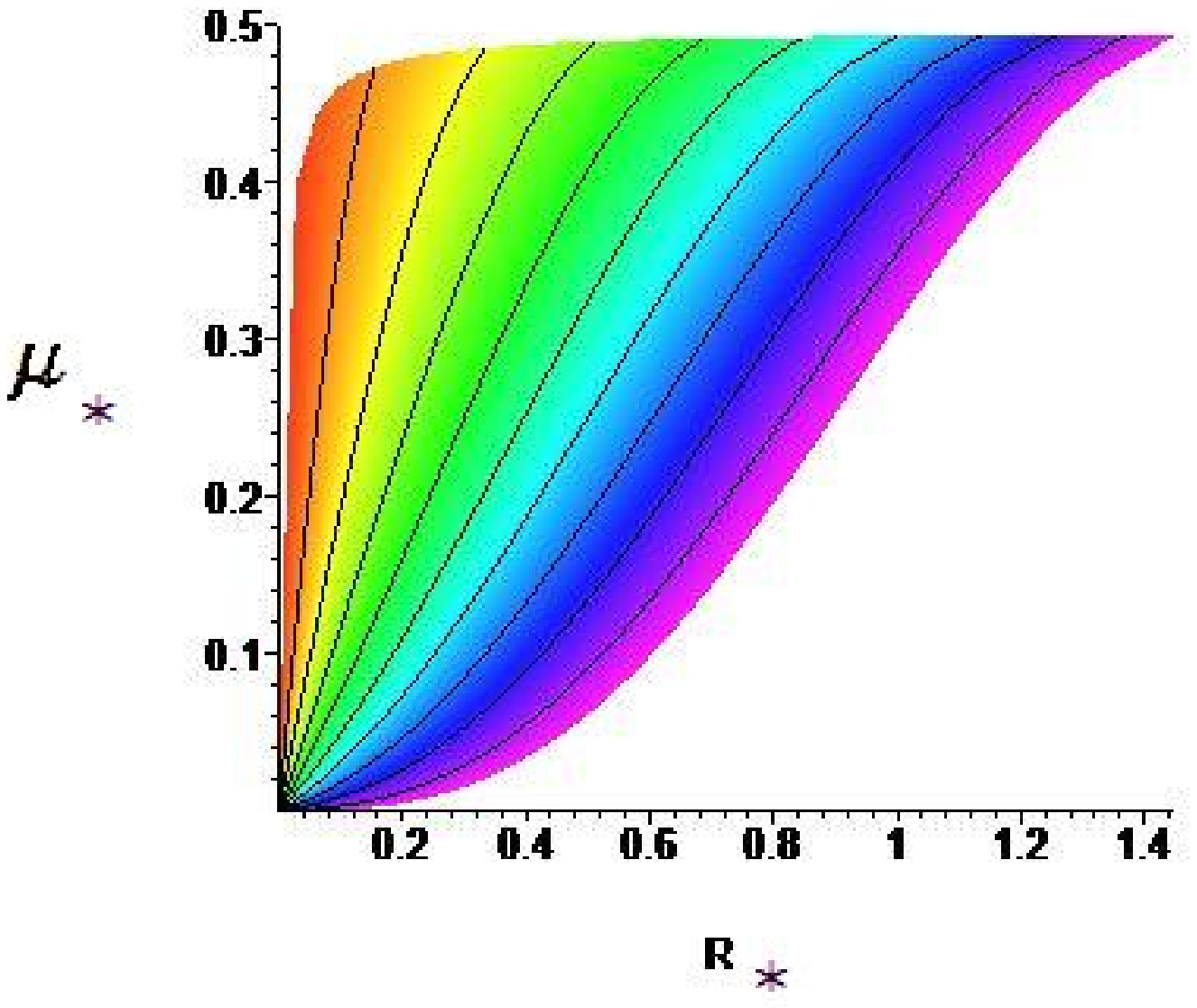} \caption{\hskip 0.2truecm The domain ${\mathbb
D}^{(2)}_{\mu_*,R_*}$ for the function $\xi_C(\mu_*;{\cal R}_*)$.
    \hskip 1truecm}
    \label{Fig8}
\end{figure}

Parts of the corresponding domains ${\mathbb D}^{(2)}_{\mu_*,R_*}$
are shown in Figs. \ref{Fig7} and \ref{Fig8}, where the condition
(\ref{Dxixi:z}) is still not taken into account.

\section{The Critical Radius and Critical Mass of Incompressible Stars}

The limiting case of Eq. (\ref{Dxixi:z}) yields the following
equation for the critical value $\xi_*^{crit}=\xi_*^{crit}(\xi_C)$
of the dimensionless luminosity variable:
\ben J_{4|3}\left(\xi_*^{crit},\xi_C\right)\!-\!
J_{4|3}\left(\xi_C,\xi_C\right)\!=\!{{2\,\xi_*^{crit}}\over
{3\sqrt{-P_4(\xi_*^{crit},\xi_C)}}}.\hskip .8truecm\la{xi_crit}
\een
%

It is obvious that this equation defines a new function
$\xi_*^{crit}(\xi_C)$, which describes the basic difference
between GR and Newtonian incompressible stars. There are no any
physical parameters in the Eq. (\ref{xi_crit}). Hence, the
function $\xi_*^{crit}(\xi_C)$ is an universal mathematical one.

In the degenerated Schwarzschild case when $\xi_C=0$ we have
$-P_4(\xi,\xi_C)=\xi^2(1-\xi^2)$ and the integral in Eq.
(\ref{xi_crit}) can be easily done. This gives (see Appendix A)
$${1\over{\sqrt{1-(\xi_*^{crit})^2}}}-1={2\over{3\sqrt{1-(\xi_*^{crit})^2}}}.$$
Hence,
\ben\xi_*^{crit}(\xi_C=0)=\sqrt{8/9}\approx
.942809.\la{xi_crit_0}\een
Thus we have obtained the exact value of the new function
$\xi_*^{crit}(\xi_C)$  at the point $\xi_C=0$. It is clear that
$\xi_*^{crit}(\xi_C=1)=1$.

The shape of function $\xi_*^{crit}(\xi_C)$ is shown in Fig.
\ref{Fig9}. Some of its basic values are given in Appendix B.
\begin{figure}[htbp] \vspace{5.8truecm} \includegraphics{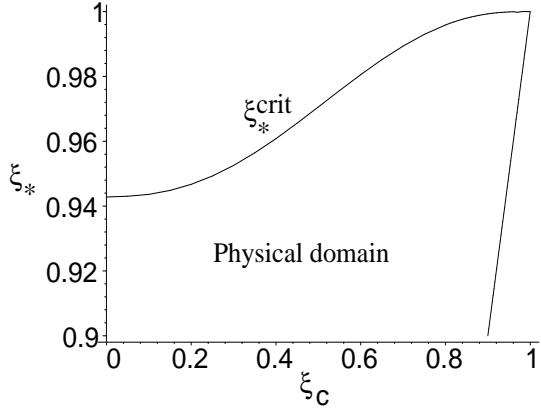} \caption{\hskip
0.2truecm The function $\xi_*^{crit}(\xi_C)$ as a border of the
physical domain of the variable
$\xi_*\in\big(\xi_C,\xi_*^{crit}(\xi_C)\big)$.
    \hskip 1truecm}
    \label{Fig9}
\end{figure}
As seen, the values of the function $\xi_*^{crit}(\xi_C)$ are in
the interval $[\sqrt{8/9},1]$. Hence, for a given value of
$\lambda$-invariant luminosity variable $\xi_C>0$, the value of
$\xi_*$ varies in the interval $[\xi_C,\xi_*^{crit}(\xi_C)]$, with
upper limit $\xi_*^{crit}(\xi_C)>\sqrt{8/9}$.

This changes radically our understanding of the relativistic
theory of stars, because now in the domain
$0\leq\xi_C\leq\xi_*\leq 1$:
\ben
{{2m_*}\over{\rho_*}}=\varsigma_*={{\xi_*^3-\xi_C^3}\over{\xi_*(1-\xi_C^3)}}\leq\xi_*^2<
\Big(\xi_*^{crit}(\xi_C)\Big)^2.\hskip .4truecm\la{m_rho_xi_s}\een

As a result of the last inequality one obtains the familiar
relativistic restriction for the Schwarzschild degenerate case:
$$\Big(2m_*/\rho_*\Big)_{\xi_C=0}=\varsigma_*{}_{\big|_{\xi_C=0}}=
\xi_*^2 < 8/9.$$
This constraint yields the following restrictions on the
luminosity variable and Keplerian mass of Schwarzschild
incompressible stars:
\ben
\rho_*{}_{\big|_{\xi_C=0}}<{1\over{\sqrt{3\pi\varepsilon}}},\,\,\,\,
m_*{}_{\big|_{\xi_C=0}}< {4\over 9
}{1\over{\sqrt{3\pi\varepsilon}}}.\la{SRestriction_m_rho_0}\een
Hence, in the relativistic model with $\rho_C=0$ we are not able
to describe stars with fixed density $\varepsilon=const$ and with
arbitrary large mass $m_*$.

In contrast, in our general geometrical model of incompressible
relativistic stars with arbitrary fixed value of luminosity
variable $\xi_C\in [0,1]$ we have the restrictions:
\ben \rho_*<\rho_*^{crit}\!:=\!{1\over{\sqrt{8\pi\varepsilon/3}}}
{{\xi_*^{crit}(\xi_C)}\over{\sqrt{1\!-\!\xi_C^3}}}
\to\infty:\,\,\,\hbox{for}\,\,\xi_C\!\to\!1\!-\!0,\nonumber\\
m_*<\!m_*^{crit}\!:=\!{1\over 2}
{1\over{\sqrt{8\pi\varepsilon/3}}}
{{\left(\xi_*^{crit}(\xi_C)\right)^3\!-\!\xi_C^3}\over
{\left(\sqrt{1\!-\!\xi_C^3}\right)^3}}
\to\infty:\hskip .7truecm\\
\,\,\,\hbox{for}\,\,\xi_C\to 1-0.\nonumber
\la{Restriction_m_rho_0}\een These upper limits
$\rho_*^{crit}(\xi_C)$ and $m_*^{crit}(\xi_C)$ are {\em exact},
i.e., the corresponding quantities $\rho_*$ and $m_*$ can become
arbitrary close to them for a given value of $\xi_C$.

\begin{figure}[htbp] \vspace{5.8truecm} \includegraphics{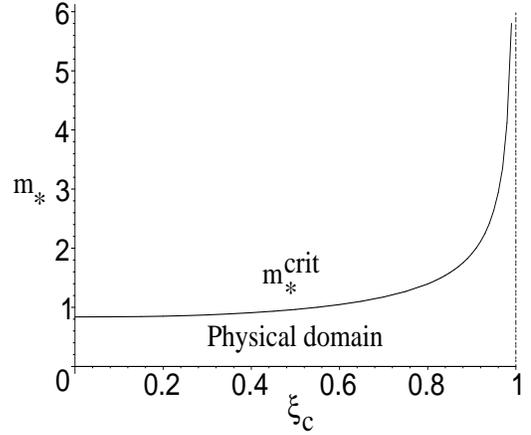} \caption{\hskip
0.2truecm The critical mass of the star  $m_*^{crit}(\xi_C)$ as a
border of the physical domain of the variable $m_*\in
\big(0,m_*^{crit}(\xi_C)\big)$.
    \hskip 1truecm}
    \label{Fig10}
\end{figure}

The calculation of the limit of $\rho_*^{crit}(\xi_C)$ is obvious
from $\xi_*^{crit}(\xi_C=1-0)=1$, see Fig. \ref{Fig9}. The
analytical proof of the above limit for $m_*^{crit}(\xi_C)$ when
$\xi_C\to 1-0$ follows immediately from our Proposition 3 (see
Section IV, C) and Eq.(\ref{mumu0}). The behavior of the critical
mass $m_*^{crit}(\xi_C)$ as a function of the central value
$\xi_C$ of luminosity variable is shown in Fig. \ref{Fig10} for
density $32\pi\varepsilon/3=1$ \footnote{For the value
$\varepsilon=3/32\pi$ the critical mass $m_*^{crit}(\xi_C)$ of
incompressible relativistic stars becomes an universal
mathematical function, like the function $\xi_*^{crit}(\xi_C)$.}.

Thus we have arrived at the following

{\bf \em Proposition 3:} {\em  For any fixed density
$\varepsilon=const$ in our essentially non-Euclidean general model
of relativistic stars there exist incompressible stars with
arbitrary large mass $m_*$.}

Obviously, stars with arbitrary large mass $m_*$ exist only for
large enough values of luminosity variable $\xi_C$.

\begin{figure}[htbp] \vspace{5.8truecm} \includegraphics{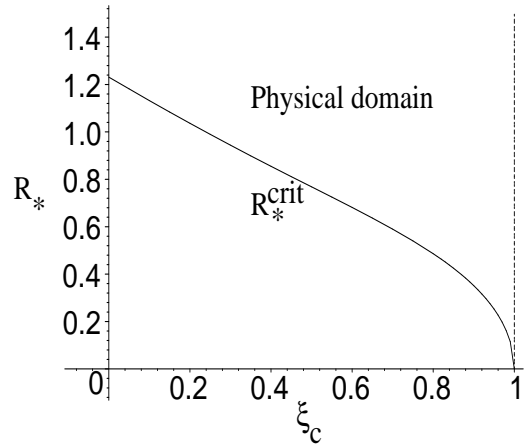} \caption{\hskip
0.2truecm The critical geometrical radius of the star
$R_*^{crit}(\xi_C)$ as a border of the physical domain of the
geometrical stelar radius
$R_*\in\big(R_*^{crit}(\xi_C),\infty\big)$.
    \hskip 1truecm}
    \label{Fig11}
\end{figure}

From Eq. (\ref{R*}) one obtains the following critical value for
the geometrical radius of our relativistic incompressible stars:
\ben
R_*^{crit}(\xi_C)\!=\!{J_{1|1}\big(\xi_*^{crit}(\xi_C),\xi_C\big)
-J_{1|1}\big(\xi_C,\xi_C\big)
\over{\sqrt{8\pi\varepsilon/3}}}.\hskip .5truecm \la{R*_crit}\een
The behavior of the critical geometrical radius
$R_*^{crit}(\xi_C)$ as a function of the central value $\xi_C$ of
luminosity variable is shown in Fig. \ref{Fig11} for density
$32\pi\varepsilon/3=1$.

Thus, we have proved the existence of new amazing regular
relativistic objects with constant matter density: They can have
an arbitrary large mass $m_*$. As a result of strong gravity,
their geometrical radius $R_*$ and luminosity $L_*\sim 1/\rho_*^2$
are arbitrary small, when the mass $m_*$ is large enough. At the
same time in presence of such objects the geometry of the
space-time is regular everywhere. Horizons of any type do not
exist.

It seems reasonable to refer to objects as to heavy black dwarfs.
As seen in Fig. \ref{Fig10}, the existence of HBD is impossible
under widely accepted extra condition $\rho_c=0$.

The critical proper mass $m_{0*}^{crit}(\xi_C)$ is bigger then
$m_{*}^{crit}(\xi_C)$. Hence, it goes to infinity in the  limit
$\xi_C \to \infty $ together with $m_{*}^{crit}(\xi_C)$.

\begin{figure}[htbp] \vspace{5.truecm} \includegraphics{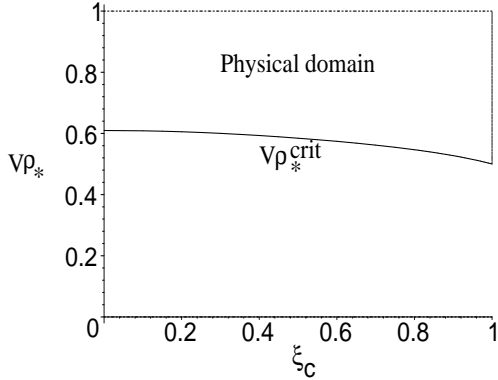} \caption{\hskip
0.2truecm The critical mass ratio $\varrho_*^{crit}(\xi_C)$ as a
border of the physical domain of the variable
$\varrho_*\in\big(\varrho_*^{crit}(\xi_C),1\big)$. (In the Fig
\ref{Fig12} the symbol $V\!\rho$ stands for $\varrho$.)
    \hskip 1truecm}
    \label{Fig12}
\end{figure}

More interesting characteristic is the $\lambda$-invariant
critical mass defect ratio $\varrho_*^{crit}(\xi_C)$, obtained
from  Eq. (\ref{varrhoxi}) in the form:
\ben \varrho_*^{crit}(\xi_C)=\hskip 5.1truecm\\
{{\big(\xi_*^{crit}(\xi_C)\big)^3-\xi_C^3}\over
{3\Big(J_{3|1}\big(\xi_*^{crit}(\xi_C),\xi_C\big)\!-
\!J_{3|1}\big(\xi_C,\xi_C\big)\Big)\sqrt{1\!-\!\xi_C^3}}} .
\nonumber\la{varrho_crit}\een
This function is represented in Fig. \ref{Fig12}.

As seen, the critical value of the mass ratio
$\varrho_*^{crit}(\xi_C)$ decreases monotonically from
$$\varrho_*^{crit}(0)={16\over9}\Bigg(1-3\sqrt{9\over8}
\arcsin\left(\sqrt{8\over9}\right)\Bigg)^{-1}\approx .609477$$ to
$1/2$ when the variable $\xi_C$ increases from $0$ to $1$.

Hence, in the case of incompressible mater the extremely strong
gravitational field, which arise in the limit $R_*\to 0$, $m_*\to
\infty$ is able to extract no more then one halve of the initial
bare mass $m_{0*}$ of the stelar matter, although the relative
part of gravitational energy increases in this limit, together
with the mass defect. It can be shown that this limitation is
EOS-dependent.

One can understand the above behavior of the critical mass ratio
$\varrho_*^{crit}(\xi_C)$ analyzing the form of the functions
$\mu_*^{crit}(\xi_C)$ -- Fig. \ref{Fig13} and
$\mu_{0*}^{crit}(\xi_C)$ -- Fig. \ref{Fig14}:

\begin{figure}[htbp] \vspace{5.truecm} \includegraphics{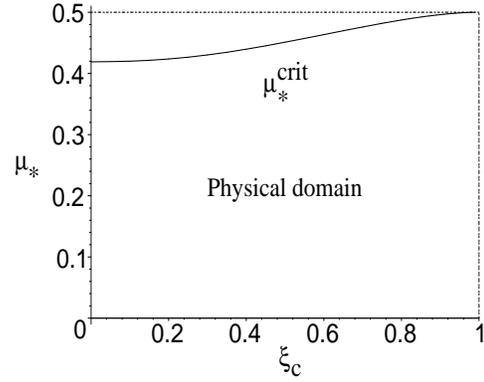} \caption{\hskip
0.2truecm The critical function $\mu_*^{crit}(\xi_C)$ as a
boundary of the physical domain of the variable
$\mu_*\in\big(0,\mu_*^{crit}(\xi_C)\big)$. As seen, on the
specific curve $\xi_*=\xi_*^{crit}(\xi_C)$ the function
$\mu_*^{crit}(\xi_C)$ goes to $1/2$, when $\xi_C\to 1$. (Compare
this specific result with Proposition 2 and with comments after
it.)
    \hskip 1truecm}
    \label{Fig13}
\end{figure}

\begin{figure}[htbp] \vspace{5.truecm} \includegraphics{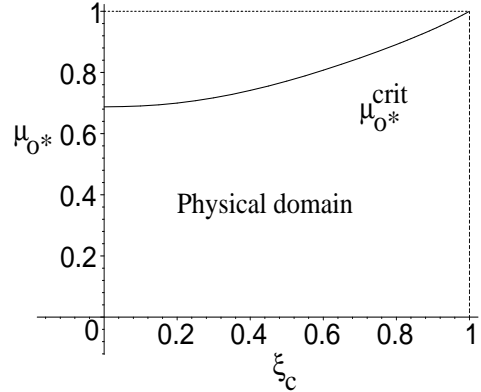} \caption{\hskip
0.2truecm The critical function $\mu_{0*}^{crit}(\xi_C)$ as a
boundary of the physical domain of the variable
$\mu_{0*}\in\big(0,\mu_{0*}^{crit}(\xi_C)\big)$.
    \hskip 1truecm}
    \label{Fig14}
\end{figure}

\begin{figure}[htbp] \vspace{7.truecm} \includegraphics{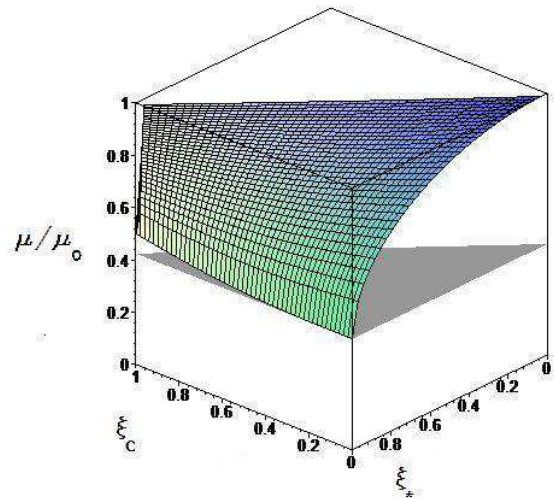} \caption{\hskip
0.2truecm The function
$\varrho(\xi_*,\xi_C)=\mu(\xi_*,\xi_C)/\mu_0(\xi_*,\xi_C)\geq
4/3\pi$ for incompressible star. The shadowed triangle is a part
of the horizontal plane $4/3\pi\approx .424413$.
    \hskip 1truecm}
    \label{Fig15}
\end{figure}

A more deep understanding of the mass defect one can obtain
looking on 3D surface of the function,
$\varrho(\xi_*,\xi_C)=\mu(\xi_*,\xi_C)/\mu_0(\xi_*,\xi_C)\geq
4/3\pi$, shown in Fig. \ref{Fig15}. The last limit from below of
the mass ratio $\varrho(\xi_*,\xi_C)$ originates from EOS for
incompressible  stars Eq. (\ref{inc}) and will be different for
relativistic stars with other EOS. It reflects the form of the
right border of the 3D surface:
\ben\varrho_*(\xi_*,0)={2\over 3}
{{\xi_*^3}\over{\arcsin\xi_*-\xi_*\sqrt{1-\xi_*^2}}},
\la{Rborder}\een
calculated for the Schwarzschild degenerate case of incompressible
stars. An analogous simple expression can be drown for
$\varrho\left(\xi_*,\sqrt[3]{1/4}\right)$, using results, given in
Appendix A.

The real limit from below on $\varrho(\xi_*,\xi_C)$ in the problem
at hands is $1/2$. It reflects the existence of the curve
$\xi_*^{crit}(\xi_C)$, which is not shown in Fig.\ref{Fig15}.

\section{Mass-Radii Relations for White Dwarfs and
         the New Geometrical Models of GRS}

Our previous consideration \cite{F04NMGS} showed that there exist
two different approaches to the GRS with stiff mass-radii
functional dependence:

1) The standard one, based on the extra condition
$\rho(0)\equiv\rho_C=0$, and

2) A new one, with extra condition: $\digamma$-- fixed. It arises
naturally in basic regular radial gauge \cite{F04NMGS}. Here
$\digamma\in(-\infty,\infty)$ is the stelar crust parameter. It
defines the jump of the derivative of luminosity variable
$\rho(r)$ as a function of the radial one, $r$ :
$\rho^\prime(r_*-0)=e^{-\digamma}\rho^\prime(r_*+0)$, at the edge
of the star $r_*$.

At the same time one can consider a geometrical models of GRS
without any additional condition of that type. In this most
general case stiff mass-radii relations, based only on GR
considerations, i.e. independent on EOS, do not exist
\cite{F04NMGS}. In this case $m_*$ and $R_*$ are free parameters,
constrained in some 2D domain, the precise form of which depends
on the EOS.

\begin{figure}[htbp] \vspace{6.truecm} \includegraphics{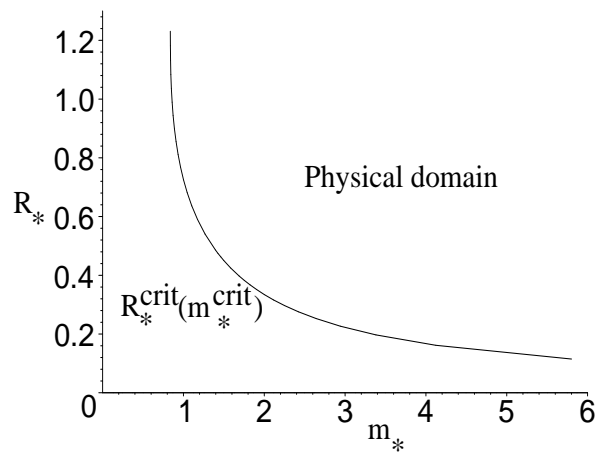} \caption{\hskip
0.2truecm The function $R_*^{crit}(m_*^{crit})$  as a boundary of
physical domain for incompressible star with energy density
${{32}\over 3}\pi\varepsilon=1$.
    \hskip 1truecm}
    \label{Fig16}
\end{figure}

The physical domain of variables $m_*$ and $R_*$ for
incompressible GRS is shown in Fig. \ref{Fig16}.
It is obvious that this physical domain is very similar to the
one, recently observed for real white dwarfs, see Fig. 2 in the
article by  J.~Madej, M.~Nale\'zyty and L.~G.~Althaus in
\cite{WD}. This demonstrates how we are able to reproduce the real
observational data in a qualitatively right way, using the
simplest new geometrical model of incompressible GRS.

The most important consequence of our analysis of the
observational data is the conclusion that actually we {\em do not}
observe stiff mass-radii relations in Nature. Instead, these data
demonstrate a clear indication that $m_*$ and $R_*$ are filling a
2D {\em domain} with a sharp boundary of the type, similar to the
one, shown in our Fig. \ref{Fig16}. This corresponds to our
general models of GRS without any extra conditions \cite{F04NMGS}.
Therefore we will skip the consideration of such relations in the
present article, although they may be of some interest, too.

For a more precise qualitative treatment of the observational data
one need to obtain new geometrical models of the GRS with
realistic EOS. This may give a basis for explanation of the mass
distributions of special type of stars, like white dwarfs, or
neutron stars, independently of theory of gravity. We will present
the corresponding results elsewhere.

\section{Concluding Remarks}

In conclusion several remarks have to be made:

1. In the present article we have recovered an interesting new GR
phenomena, studying the most general solutions of ETOV equations
for incompressible matter. We were able to find proper physical
meaning of {\em all} solutions of these equations, overcoming the
requirement for regularity of solutions at the zero value of the
luminosity variable $\rho$. As we have seen, this way we obtain a
two parameters family of solutions.

2. We wish to comment once more the most important result of
present article: the proof of the existence of new GR objects --
heavy black dwarfs (HBD).

From physical point of view this becomes possible, because the
fulfilment of the condition $\rho \geq \rho_C > \rho_G$
\cite{F04NMGS}, where $\rho_G=2 m_*$ is the Schwarzschild radius,
is equivalent to the introduction of non-permeable potential wall
in the space of the luminosity variable $\rho$. Because of the
Gauss theorem, this introduces  in the energy balance of the body
an effective "potential energy" with an infinite jump.

Then, following the original Landau energetic consideration of the
problem \cite{Landau}, we conclude that the presence of any
additional energy terms, due to EOS, can not destroy the existence
of static equilibrium of the body, at least in the case of
non-ultra-relativistic matter. This "$\rho$-space picture"
explains in a more usual language the consequences of the new
boundary conditions at the real center of the star, under which we
are solving Einstein equations. See for details \cite{F04NMGS}.

In terms of the proper radial variable $r$ no wall exist at all.
Thus we see once more, that one is to give up the idea to consider
the luminosity variable $\rho$ as a measure of distances. It is
responsible only for determination of geometrical area and
luminosity of physical objects.

3. The results of the present article raise many new questions and
problems.

For example, we obtain a new real alternative to the black holes
(BH) for explanation of nature of the massive compact dark
objects, observed in astrophysics. Hence, we need a new criteria
to distinguish BH and HBD.

An obvious problem is the possible relation of HBD with the
observed could dark matter in the universe, as well as the
relation of HBD with gravitational collapse of usual bodies. It is
clear, that the present series of articles calls for a novel
consideration of the collapse, too.

For a comparison with observational data, we obviously need
consideration of new geometrical models of GRS with more realistic
EOS for stelar matter.

We intend to address all these questions in the forthcoming
articles in this series.


 \vskip 1.truecm

{\em \bf Acknowledgments} \vskip .3truecm

The author is grateful to the JINR, Dubna for financial support of
the present article and for the hospitality and good working
conditions during his two three-months visits in 2003 and in 2004,
when the most of the work has been done.

The author is deeply indebted to Prof. T.~L.~Boyadjiev for
friendly support and many useful discussions on mathematical
problems, related to the stelar boundary problem, to Prof.
S.~Bonazzola, for discussions, and especially, for encouraging
general information about the existence of results in direction of
overcoming of the standard relativistic restriction on the stelar
masses. Similar results can  be found in the articles \cite{Bell},
too. The author wishes to thank Prof. Ll.~Bell for attracting his
attention to these articles.

The author is grateful, too, to E.-M. Pauli and to M. Miller for
discussions on white dwarfs' physics and for information about
available data. Special tanks to E.-M. Pauli for sending the file
of her PhD thesis.

\appendix

\section{Calculation of the Elliptic Integrals}

Using standard incomplete elliptic integrals of first, second and
third kind ($F$, $E$ and $\Pi$) \cite{elliptic}, we can write down
the integrals
\ben J_{n|m}(\xi,\xi_C):=\int\!d\xi\,{{\xi^n
}\Big/\left({\sqrt{-P_4(\xi,\xi_C)}}\right)^m},\la{Jnm}\een
which are needed for the relativistic incompressible star problem,
in the following form:

I. $n=0$, $m=1$:
\ben J_{0|1}(\xi,\xi_C)= g\,{1\over i}F(iz,k),\la{J01}\een\\
In the generate cases one obtains:\\

\noindent a)\,\, $J_{0|1}(\xi,0)={1\over 2}
\ln\left({{1-\sqrt{1-\xi^2}}\over{1+\sqrt{1-\xi^2}}}\right);$\\

\noindent b)\,\,$J_{0|1}(\xi,\sqrt[3]{1/4})= {2\over {\sqrt{3}}}
\arctan
\left({{\sqrt{3}(-1+4\xi)}\over{6\sqrt{\xi(1-\xi)}}}\right).$\\

II. $n=1$, $m=1$:
\ben J_{1|1}(\xi,\xi_C)= g\,{1\over i}F(iz,k)+g\,{1\over
i}\Pi(iz,\alpha^2,k),\la{J11}\een\\
In the generate cases one obtains:\\

\noindent a)\,\, $J_{1|1}(\xi,0)=\arcsin(\xi);$\\

\noindent b)\,\,$J_{1|1}(\xi,\sqrt[3]{1/4})\!=\!
{{1}\over{3}}\sqrt{3}\arctan\left(\!
{{\sqrt{3}(1-4\xi)}\over{6\sqrt{\xi(1-\xi)}}}\!
\right)\!-\!\arcsin(1\!-\!2\xi)$.\\

II. $n=2$, $m=1$:
\ben J_{2|1}(\xi,\xi_C)= {{1-3\sigma}\over 4}g\,{1\over
i}F(iz,k)+\nonumber\\{2\over g}\,{1\over
i}E(iz,k)-{{\sqrt{-P_4(\xi,\xi_C)}}\over{1-\xi}},\la{J21}\een
In the generate cases one obtains:\\

\noindent a)\,\, $J_{2|1}(\xi,0)=1-\sqrt{1-\xi^2};$\\

\noindent b)\,\,$J_{2|1}(\xi,\sqrt[3]{1/4})= -\sqrt{\xi(1-\xi)}
-{{\sqrt{3}}\over{6}} \arctan\left(\!
{{\sqrt{3}(1-4\xi)}\over{6\sqrt{\xi(1-\xi)}}}\!\right).$\\

II. $n=3$, $m=1$:
\ben J_{3|1}(\xi,\xi_C)= {{2-\xi_C^3}\over 4}\,g\,{1\over
i}F(iz,k)-\nonumber\\{{1-\xi_C^3}\over 2}\,g\,{1\over
i}\Pi(iz,\alpha^2,k)-{1\over 2}
\sqrt{-P_4(\xi,\xi_C)},\la{J31}\een\\
In the generate cases one obtains:\\

\noindent a)\,\, $J_{3|1}(\xi,0)={1\over 2}\arcsin(\xi)-
{1\over 2}\xi\sqrt{1-\xi^2};$\\

\noindent b)\,\,$J_{3|1}(\xi,\sqrt[3]{1/4})=
{{\sqrt{3}}\over{12}}\arctan\left(\!
{{\sqrt{3}(1-4\xi)}\over{6\sqrt{\xi(1-\xi)}}}\!\right) -\\
\hbox{\hskip 1.5truecm}{3\over 8}\arcsin(1-2*\xi)
-{1\over 4}(1+2\xi)\sqrt{\xi(1-\xi)}.$\\

II. $n=4$, $m=3$:
\ben J_{4|3}(\xi,\xi_C)\!=\!{{(1+\sigma)(\sigma^2-6\sigma-3)}
\over{\sigma(3+\sigma)(9-\sigma^2)}}\,g\,{1\over
i}F(iz,k) + \nonumber\\
\left({{2(3+\sigma^2)}\over{\sigma(9-\sigma^2)}}\right)^2{2\over
g}\,{1\over
i}E(iz,k)- \nonumber\\
{{\xi\left(2(3+\sigma^2)\xi+(1-\sigma^2)(3-\sigma^2)\right)}
\over{\sigma^2(9-\sigma^2)\sqrt{-P_4(\xi,\xi_C)}}}\, .\la{J43}\een

In the generate cases one obtains:\\

\noindent a)\,\, $J_{4|3}(\xi,0)=1/\sqrt{1-\xi^2}-1;$\\

\noindent b)\,\,$J_{4|3}(\xi,\sqrt[3]{1/4})={5\over{27}}\sqrt{3}
\arctan{\left({{\sqrt{3}(1-4\xi)}\over{6\sqrt{\xi(1-\xi)}}}\right)}+
\\
\hbox{\hskip 3.truecm}{1\over
18}{{4\xi^2+15\xi+5}\over{(\xi-1/2})^2}\sqrt{{{\xi}\over{1-\xi}}}
-{{5\pi\sqrt{3}}\over{54}}$.\\

Here $\sigma:=\sqrt{1-4\xi_C^3}$,\,\,\,
$z:=\alpha^{-1}\sqrt{\xi/(1-\xi)}$,
\ben
g:={4\over{\sqrt{(1+\sigma)(3-\sigma)}}},\nonumber\\
k:=\sqrt{{{1-\sigma}\over{1+\sigma}}\,
{{3+\sigma}\over{3-\sigma}}},\nonumber\\
\alpha:=\sqrt{{{1-\sigma}\over{3-\sigma}}},
\,\,\,\hbox{and}\la{parameters}\een
\begin{subequations}\label{integrals:abc}
\ben {1\over i}F(iz,k):=\int_0^z
{{dx}\over{\sqrt{(1+x^2)(1+k^2x^2)}}},
\hskip 1.5truecm\label{integrals:a}\\
{1\over i}E(iz,k):=\int_0^z\sqrt{ {{1+k^2x^2}\over{1+x^2}}} dx,
\hskip 2.5truecm\label{integrals:b}\\
{1\over i}\Pi(iz,\alpha^2,k):=\!\int_0^z\!\!
{{dx}\over{\!(1\!+\!\alpha^2x^2)\!\sqrt{\!(1\!+\!x^2)(1\!+\!k^2x^2)}}\!}
\hskip .3truecm\label{integrals:c} \een \end{subequations}
define for any values $0\!\leq\!\xi_C\!\leq\!\xi\!<1$ three basic
elliptic integrals in most convenient for us uniform
representation.

Note that for $\xi_C\in\big[0,\sqrt[3]{1/4}\,\big]$ the parameters
(\ref{parameters}) have real values: $k\in \big[0,1\big]$, $g\in
\big[2,4/\sqrt{3}\big]$, $\alpha\in \big[0,1/\sqrt{3}\big]$.

For $\xi_C\in\big(\sqrt[3]{1/4},1\big)$ these parameters are
complex numbers. Nevertheless, for $0\leq\xi_C\leq\xi<1$ the
integrals (\ref{J01})-(\ref{J43}) have real values in this case,
too.

Of course one can use more complicated representations of the
standard elliptic integrals \cite{elliptic}, such that the
integrals $J_{n|m}(\xi,\xi_C)$ will have transparently real values
for $\xi_C\in\big(\sqrt[3]{1/4},1\big)$:

Taking into account that in this case $k=\exp(i\psi)$, $\psi\in
\Re$, we can perform the transformation of the modulus $k\to\tilde
k={1\over 2}\left(\sqrt{k}+1/\sqrt{k}\right)=\cos(\psi/2)\in[0,1]$
in all elliptic integrals. For example, $F(z,k)=F(\tilde z,\tilde
k)/2\sqrt{k}$, where $\tilde z^{-1}={1\over
2}\left(z\sqrt{k}+1/z\sqrt{k}\right)$. Unfortunately, such
representations for $E(z,k)$ and $\Pi(z,\alpha,k)$ are quite
complicated. We will not give them here. They can be obtained
composing several transformations of $k$, described in
\cite{elliptic}.


\section{Table of the values of function $\xi_*^{crit}(\xi_C)$}
\begin{table}[here]~\vspace{-1.truecm}
\caption{The values of function $\xi_*^{crit}(\xi_C)$}
\begin{center}
\begin{tabular}{cllllrrrr}
\hline \hline
$\xi_C$        & $\Big|$ & 0.00  & 0.05  & 0.10  & 0.15  & 0.20  & 0.25  & 0.30 \\
\hline
$\xi_*^{crit}$ & $\Big|$ & .9428 & .9430 & .9436 & .9448 & .9467 & .9493 & .9525 \\
\hline \hline
$\xi_C$        & $\Big|$ & 0.35  & 0.40  & 0.45  & 0.50  & 0.55  & 0.60  & 0.65 \\
\hline
$\xi_*^{crit}$ & $\Big|$ & .9564 & .9608 & .9 656 & .9706 & .9756 & .9806 & .9852 \\
\hline \hline
$\xi_C$        & $\Big|$ & 0.70  & 0.75  & 0.80  & 0.85  & 0.90  & 0.95  & 1.00 \\
\hline
$\xi_*^{crit}$ & $\Big|$ & .9894 & .9930 & .9959 & .9980 & .9993 & .9999 & 1.00 \\
\hline \hline
\end{tabular}
\end{center}
\end{table}

\vskip 0truecm

\end{document}